\documentclass[aip,reprint,twocolumn]{revtex4-2}

\usepackage{graphicx}
\usepackage{dcolumn}
\usepackage{bm}
\usepackage{color}
\usepackage{amsmath}
\usepackage{amssymb}
\usepackage{graphicx}
\usepackage[unicode=true,pdfusetitle,
 bookmarks=true,bookmarksnumbered=false,bookmarksopen=false,
 breaklinks=false,pdfborder={0 0 0},pdfborderstyle={},backref=false,colorlinks=true]
 {hyperref}
\hypersetup{
 pdfborderstyle={},citecolor=blue}
\usepackage[utf8]{inputenc}
\usepackage[T1]{fontenc}
\usepackage{mathptmx}

\begin{document}

\preprint{AIP/123-QED}

\title[Square Waves and Bykov T-points]{Square Waves and Bykov T-points in a Delay Algebraic Model for the Kerr-Gires-Tournois Interferometer}
\author{Mina Stöhr$^{\dagger}$}
\affiliation{Weierstrass Institute, Mohrenstrasse 39, 10117 Berlin, Germany}
\author{Elias R. Koch$^{\dagger}$}%
\affiliation{Institute for Theoretical Physics, University of M\"unster, Wilhelm-Klemm-Str. 9, 48149 M\"unster, Germany}
\author{Julien Javaloyes}
\affiliation{Departament de Física \& IAC-3, Universitat de les Illes Balears, C/ Valldemossa km 7.5, 07122 Mallorca, Spain}
\author{Svetlana V. Gurevich}
\affiliation{Institute for Theoretical Physics, University of M\"unster, Wilhelm-Klemm-Str. 9, 48149 M\"unster, Germany}
\affiliation{Center for Nonlinear Science (CeNoS), University of M\"unster, Corrensstrasse 2, 48149 M\"unster, Germany}
\author{Matthias Wolfrum}
\affiliation{Weierstrass Institute, Mohrenstrasse 39, 10117 Berlin, Germany}%

\date{\today}

\begin{abstract}
We study theoretically the mechanisms of square wave
formation of a vertically emitting micro-cavity operated in the Gires-Tournois regime that contains a Kerr medium and that is subjected to strong time-delayed optical feedback and detuned optical injection. We show that in the limit of large delay, square wave solutions of the time-delayed system can be treated as relative homoclinic solutions of an equation with an advanced argument. Based on this, we use concepts of classical homoclinic bifurcation theory to study different types of square wave solutions. In particular, we unveil the mechanisms behind the collapsed snaking scenario of square waves and explain the formation of complex-shaped multistable square wave solutions through a Bykov T-point. Finally we relate the position of the T-point to the position of the Maxwell point in the original time-delayed system.
\end{abstract}

\maketitle

\begin{quotation}

The formation of complex  patterns can be observed in a huge variety of nonlinear systems. Usually, they are described as spatial patterns in systems of partial differential equations. However, also in certain time delay systems, for example modeling optical and optoelectronic systems with a feedback loop, similar phenomena can be observed in a purely temporal evolution without any spatial variables. This analogy of time-delay systems in the regime of large delay and spatially extended system has been discussed in many aspects and dynamical phenomena as for example  solitons and localized structures, weakly interacting pulses, or coexisting wave patterns have been studied. Here, we present another example of this analogy: a delay-algebraic equation modeling a vertical external-cavity Kerr-Gires-Tournois interferometers in presence of anti-resonant injection. In this system we observe the formation of complex coexisting square-wave patterns that are organized in a structure of snaking branches around a Maxwell point. We will elaborate a methodology how they can be systematically studied. As in the case of similar spatial patterns, it is based on homoclinic bifurcation theory and numerical path following methods. 
\end{quotation}

\section{\label{sec: Introduction} Introduction}
Real-world complex systems can be strongly influenced by time delays due to unavoidable finite signal propagation speeds and time-delayed systems have proven to be a fertile framework for the modeling of the resulting nonlinear dynamical phenomena~\cite{YanchukGiacomelli2017spatiotemporal,EJW-CHA-17}. A typical delay induced dynamical phenomenon is the formation of square wave (SW) oscillations if the delay time is sufficiently large. These are periodic solutions consisting of sharp transitions between two (or more) alternating plateaus. SWs have been studied extensively from a mathematical point of view \cite{MalletParet1986GlobalCA,HaleHuang1994PeriodDoubling,HaleHuang1996singularlyperturbedDDE,N-PD-03} and also found applications such as optical clocks in signal processing, communication systems~\cite{KAA-PTL-10,SWX-OL-13} or optical sensing~\cite{USN-OE-11}. They have been observed experimentally and analyzed theoretically in several optical and opto-electronic systems, e.g., vertical-cavity surface-emitting lasers \cite{MGJ-PRA-07,MJG-PRA-13}, edge-emitting diode lasers \cite{GES-OL-06,Friart:14}, semiconductor ring lasers \cite{LLZ-OL-16,Mashal:12}, and quantum dot lasers\cite{Dillane:19}, to name just a few. 
In the framework of time-delayed systems, SWs typically appear in the regime of large delay via a supercritical Andronov-Hopf (AH) bifurcation with a 50\% duty cycle and a period of approximately twice the delay \cite{N-PRE-04,Mashal:12}. Additionally, asymmetrical SWs as well as SWs with a period close to once the delay have been observed in e.g., a broadband bandpass optoelectronic oscillator~\cite{WED-PRE-12}.

However, another mechanism of  SW formation was reported recently in nonlinear vertical external-cavity-Gires-Tournois interferometers~\cite{GT-CRA-64} enclosed into a long external feedback cavity in the presence of anti-resonant injection. There, it was demonstrated that in the normal dispersion regime SWs can appear via a collapsed snaking scenario~\cite{KW-PhysD-05,BurkeKnobloch2007HomoclinicSnaking}, when either Kerr~\cite{KSG-OL-22} or semiconductor quantum well~\cite{KSJG-23} nonlinearities are employed. This leads to the formation of complex-shaped multistable SW solutions. Indeed, due to the oscillatory tails induced by the dispersive microcavity~\cite{SCM-PRL-19}, moving fronts between different plateau solutions can lock at several positions in the vicinity of the so-called Maxwell point, where the two fronts have the same speed and their dynamics is arrested. In the same optical system but with resonant optical feedback, dark and bright temporal localized states (TLSs) are shown to appear via the locking of domain walls between bistable continuous wave background states~\cite{SPV-OL-19,SJG-OL-22,SGJ-PRL-22}. TLSs possess a period close to the round-trip in the external cavity, and, like in the case of SWs, they can interlock at multiple equilibrium distances resulting in a collapsed snaking bifurcation scenario. Note that the collapsed snaking has been observed in many different physical systems, ranging from the dynamics of thin liquid films and flame propagation to the vegetation patterns and optical pulses,  see e.g.~\cite{YBK-PRL-06,TGT-EPJE-14,JBK-CF-17,PF-PRE-20,PAM-PRA-21,GomilaKnobloch2021HomoclinicSnaking}. 

Recently, it was shown that in the framework of time-delayed systems and in the long delay limit, TLSs can be treated as homoclinic solutions of an equation with an advanced argument which determines the TLS profile~\cite{YanchukRuschelSieberWolfrum2019TDS}. This so-called profile equation allowed to apply the tools of classical homoclinic bifurcation theory~\cite{HomburgSandstede2010HomoclinicBifurcations} to study different types of solutions, their bifurcations and instabilities~\cite{StohrWolfrum2023TDS}. 
In this paper we extend the homoclinic bifurcation theory tools introduced in \cite{YanchukRuschelSieberWolfrum2019TDS,StohrWolfrum2023TDS} to the SW solutions and apply them to a time-delayed model of Kerr-Gires-Tournois (KGTI) interferometers. Using a combination of analytical, numerical and path-continuation methods, we demonstrate that SWs can be treated as relative homoclinic orbits with respect to a mirror symmetry. Further, we shall show that the snaking scenario and the corresponding Maxwell point can be analyzed as a Bykov T-point~\cite{GlendinningSparrow1986Tpoints,Bykov1993Tpts,KnoblochLamb2014Bykov}.  This is a bifurcation of homoclinic orbits and can also be tracked in parameter space using numerical path-continuation techniques~\cite{StohrWolfrum2023TDS}.

\section{\label{sec: SW in KGTI}Model Equations}
 \begin{figure}
	\includegraphics[width=1\columnwidth]{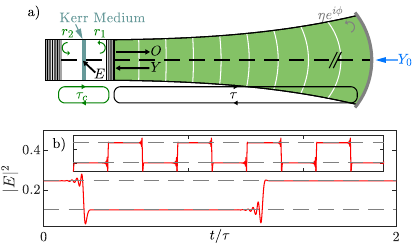}
	\caption{ (a) Schematic of a micro-cavity containing a Kerr medium coupled to a long external cavity which is closed by a mirror with reflectivity $\eta$ and phase $\phi$, and driven by a CW beam with amplitude $Y_0$.  (b) An exemplary SW solution obtained from a direct numerical simulation of Eqs.~\eqref{eq: KGTI1}, \eqref{eq: KGTI2}. The gray dashed lines correspond to the period-two orbit of the singular map \eqref{eq: singular map}. The inset shows the dynamics over several round-trips. 
 } \label{fig:fig1}
 \end{figure}
The schematic setup is depicted in Fig.~\ref{fig:fig1}~(a), see also~\cite{SPV-OL-19,SJG-OL-22,KSG-OL-22} for more details. It is composed of a monomode micro-cavity with round-trip time $\tau_c$. The micro-cavity is of a few micrometers in length and has a
radius up to 100 $\mu$m. It contains a thin layer of the Kerr material acting as a nonlinear medium that is situated at the anti-node of the field. The micro-cavity is closed by two distributed Bragg
mirrors with reflectivities $r_{1,2}$, whereas the long external cavity of a few centimeters and round trip time $\tau\gg\tau_c$ is closed by a mirror with reflectivity $\eta$ and the feedback phase $\phi$. The system is driven by continious wave (CW) injection with amplitude $Y_0$ and frequency $\omega_0$. The total external cavity phase $\varphi=\phi+\omega_0\tau$ is the sum of the propagation phase in the external cavity and of the phase shift induced by the feedback mirror. It describes the detuning with respect to the nearest external cavity mode and $\varphi=\pi$ corresponds to the situation where the injection frequency is set exactly inbetween two external cavity modes. In this case trains of SWs with periodicity $\gtrsim2\tau$, as shown in the inset of Fig.~\ref{fig:fig1}~(b), can be generated for a range of $Y_0$ and of the detuning $\delta=\omega_c-\omega_0$, where $\omega_c$ is the micro-cavity resonance.

To analyze the dynamics of the SWs of the system depicted in Fig.~\ref{fig:fig1}, we employ the first-principles model obtained 
by solving exactly the field equations in the linear parts of
the micro-cavity and connecting the fields at the interface with the
nonlinear medium as a boundary condition~\cite{MB-JQE-05}. Then the evolution of the (normalized) slowly varying field envelopes in the micro-cavity $E$ and the external cavity $Y$ is governed by~\cite{SPV-OL-19,SJG-OL-22,KSG-OL-22}
\begin{align}
\dot{E}&=\left[i\left(|E|^{2}-\delta\right)-1\right]E+hY, \label{eq: KGTI1}\\
Y&=\eta e^{i\varphi}\left(E\left(t-\tau\right)-Y\left(t-\tau\right)\right)+\sqrt{1-\eta^{2}}Y_{0}.
\label{eq: KGTI2}
\end{align}
Here, the coupling between the intra- and external cavity fields is given by a delay algebraic equation (DAE)~\eqref{eq: KGTI2} which takes into account all the multiple reflections in the external cavity. Further, $h=h(r_1,r_2)=(1+|r_2|)(1-|r_1|)/(1-|r_1||r_2|)$ is the light coupling efficiency. Note that for a perfectly reflective bottom mirror one obtains $h(r_1,1)=2$ corresponding to the so-called Gires-Tournois interferometer regime \cite{GT-CRA-64}. Hence, second- and third-order dispersions are naturally captured by Eqs.~\eqref{eq: KGTI1}, \eqref{eq: KGTI2}. Notice, that due to third-order dispersion becoming the leading term around resonance, the resulting SW solutions can possess strong oscillatory tails, see Fig.~\ref{fig:fig1}~(b), where the typical time trace of an SW obtained from numerically integrating Eqs.~\eqref{eq: KGTI1}, \eqref{eq: KGTI2} is presented.

\begin{figure}
	\includegraphics[width=1\columnwidth]{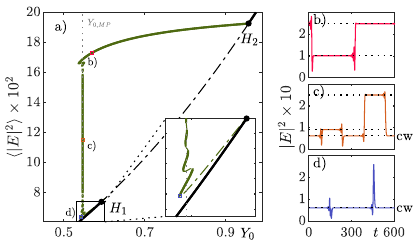}
	\caption{ (a) Branch of periodic solutions (green) and CW states (black) for varying $Y_0$ obtained by path-continuation of Eqs.~\eqref{eq: KGTI1}, \eqref{eq: KGTI2}. Stable (unstable) solutions are depicted in solid (dashed), respectively. $H_{1,2}$ are a subcritical and a supercritical AH-bifurcation of the CW state. The branch consists of SW solutions (b) in the upper part, mixed-type solutions (c) around the Maxwell point $Y_{0,MP}\approx 0.54625$ (dotted grey), and unstable alternating pulse solutions (d) close to the subcritical AH-bifurcation point $H_1$.  Exemplary profiles (b)--(d) are marked in (a). Parameters are $(\tau,\delta,h,\eta,\varphi)=(300,0.2,2,0.9,\pi)$.} \label{fig:fig2}
 \end{figure}
We used a recently developed extension of DDE-BIFTOOL~\cite{DDEBT,Sieber2014DDEbiftoolmanual} that allows for the bifurcation analysis of delay-algebraic systems and delay equations of neutral type. We present in Fig.~\ref{fig:fig2}~(a) a resulting bifurcation diagram  for varying $Y_0$ and other  parameters as in Fig.~\ref{fig:fig1}~(b), cf. also Ref.~\onlinecite{KSG-OL-22}. The exemplary profiles of the solutions along the branch are presented in panels (b)--(d), respectively (the solution in panel (b) is the same as in Fig.~\ref{fig:fig1}~(b)).
We notice that for the high injection values, a branch of periodic solutions (green) emerges from the CW state (black) at a supercritical AH-bifurcation point $H_2$.
The resulting upper arc of the branch consists of stable SWs, see Fig.~\ref{fig:fig2}~(b). Then, in the regime of bistability between the SWs and the stable CW state, it enters into a region of collapsed snaking around a vertical Maxwell point $Y_0=Y_{0,MP}$, experiencing a sequence of fold bifurcations, before it terminates in a subcritical AH-bifurcation point $H_1$, see the inset in Fig.~\ref{fig:fig2}~(a). Note that in the snaking region around the Maxwell point, there are multiple coexisting stable solutions consisting of fronts not only between two SW plateaus but also fronts connecting the CW state with one of the plateaus, creating stable mixed type solutions as shown in Fig.~\ref{fig:fig2}~(c). On the lower part of the branch, the SWs have transformed into unstable alternating pulse solutions hosting two localized pulses on the background of the CW state, cf. Fig.~\ref{fig:fig2}~(d). 

\section{\label{sec: SW as TDS}Treating square waves as localized states}

Recently, Yanchuk et al.~\cite{YanchukRuschelSieberWolfrum2019TDS} introduced a new approach to analyze TLSs in time-delayed systems in the long delay limit. We start our analysis by first recalling 
some general concepts of this method, and then show how it can be extended to the case of SWs, using the approach from the earlier work by Mallet-Paret and Nussbaum~\cite{MalletParet1986GlobalCA}.
For a time-delayed system given by a general  delay differential equation of the form
\begin{align}
    \dot{x}(t) = f\left(x(t),\, x(t-\tau)\right) , ~x \in \mathbf{R}^n
    \label{eq: DDE}
\end{align}
with large delay $\tau$, a TLS is a time periodic solution that can be found for all sufficiently large values of the delay $\tau$ and has a period $T(\tau)$,
which is slightly larger than the delay.  Hence, we can define a response time  
\begin{align}
    \rho(\tau):=T(\tau)-\tau,
    \label{eq: TDS rho}
\end{align}
which is positive and remains bounded, while both, $\tau$ and $T(\tau)$, can become arbitrarily large.
Such TLSs are localized in time because they spend most of the time close to a stable equilibrium solution, called {\it background}, from which they deviate only during a fixed short time interval. Equilibria that are stable for all positive values of the delay are called {\it absolutely stable}, see Ref.~\onlinecite{YanchukWolfrum2021AbsoluteStability}.

We employ now the reappearance property~\cite{YanchukPerlikowski2009DelayPeriodicity} of periodic solutions of delay equations: Any $T$-periodic solution $x_\ast (t)$ of Eq.~\eqref{eq: DDE} at delay $\tau$ is also a solution of Eq.~\eqref{eq: DDE} replacing $\tau$ by
\begin{align*}
\tau_k := \tau + kT, ~\text{for all } k \in \mathbf{Z},
\end{align*}
i.e., periodic solutions reappear when adding integer multiples of their period to the delay.
In this way, the profiles of the TLS solutions can be found as periodic solutions of Eq.~\eqref{eq: DDE} with $\tau$ replaced by $\tau_{-1}=-\rho$. 
This equation, referred to as \textit{profile equation}, is a differential equation with an advanced argument and can be solved as an initial value problem only in backward time. 
In this equation, the large delay $\tau$ has disappeared and the whole family of large-period periodic solutions can be found at finite values of $\rho$ with a limit  $\rho(\tau) \rightarrow \rho_\infty$ for $\tau \rightarrow \infty$. At $\rho=\rho_\infty$, the family of large-period periodic orbits ends in a homoclinic orbit, i.e., a trajectory $x_h (t)$, $t\in \mathbb{R}$ with 
\begin{align}\lim_{t\rightarrow \pm\infty}x_h(t)= x_0,\label{eq:hom}\end{align}
where $x_0$ is background equilibrium, which is assumed to be absolutely stable in the original equation~\eqref{eq: DDE}. Note that the profiles are exactly preserved under the reappearance map, but their stability is changed. In particular, the stable background equilibrium, that is trivially reappearing for all positive and negative delays, turns for the negative delay  $\tau=-\rho$ into an equilibrium of saddle type, such that it can serve as the limiting equilibrium (\ref{eq:hom}) of a homoclinic orbit  in the profile equation. In this way, localized solutions of delay-differential equations with large delay can be studied as homoclinic solutions of an advanced equation where the large delay has disappeared and the newly introduced advanced time shift $\rho$ corresponds to the response time (\ref{eq: TDS rho}) and remains small.

Now we show how this method can be extended to SW solutions. First, note that by
rescaling time in Eq.~\eqref{eq: DDE} by $t^\prime =t/\tau$ the resulting equation 
$$
\frac{1}{\tau} \frac{dx(t^\prime)}{dt^\prime} = f\left(x(t^\prime), x(t^\prime-1)\right) 
$$
in formal limit $\tau \rightarrow \infty$  provides 
\begin{align}
    0 = f (x(t^\prime) , x(t^\prime-1) )=:f (x_{k+1} , x_k ),
    \label{eq: singular map}
\end{align}
which can be interpreted as an implicitly given iterated map for discrete time $k\in\mathbb{Z}$ determining $x_{k+1}$ as a function of $x_k$, called {\it singular map}.
A stable fixed point $x_0$ of Eq.~\eqref{eq: singular map} is also a stable equilibrium of Eq.~\eqref{eq: DDE} and can serve as a background for TLSs. 
Suppose the equilibrium solution $x_0$ undergoes a flip bifurcation and a stable period-two solution $(x_1,x_2)$ satisfying
\begin{align}
0 = f (x_1 , x_2 ) = f (x_2 , x_1 )
\end{align}
is created. Then, these two values $x_{1,2}$ can serve as plateaus of an SW of Eq.~\eqref{eq: DDE} (cf. dashed gray lines in Fig.~\ref{fig:fig1}~(b) and Fig.~\ref{fig:fig2}~(b)-(d)), i.e., a periodic solution $x_\ast(t)$ with a period $T$ slightly bigger than $2\tau$ and two plateaus $x_{1,2}$ of length approximately $\tau$ each. 
Note that for the system
\begin{align}
    \dot{x}(t) = f (x(t), y(t -\sigma))\,,     \label{eq: APE1}\\
    \dot{y}(t) = f (y(t), x(t -\sigma))     \label{eq: APE2}
\end{align}
 the period-two orbit $(x_1,x_2)$  of the singular map provides two different equilibria $(x,y)=(x_1 , x_2 )$ and  $(x,y)=(x_2 , x_1 )$. Moreover, a straightforward calculation shows that also the SW solution $x_\ast (t)$ of Eq.~\eqref{eq: DDE} provides by  $$(x(t), y(t)):= (x_\ast (t),x_\ast (t-T/2))$$
a solution of Eqs.~\eqref{eq: APE1}, \eqref{eq: APE2} whenever
 \begin{align*}
    \sigma = \sigma_k := \tau +\left(k-\frac{1}{2}\right) T, ~\text{for all } k \in \mathbf{Z}\,.
    \end{align*}
Hence, using  $\sigma = \sigma_0 =: -\rho$, we can find the profile of the SW $x_\ast (t)$ as a solution of system~\eqref{eq: APE1}, \eqref{eq: APE2} with an advanced argument. Note that the definition~\eqref{eq: TDS rho} of the response time, which we introduced above, has to be replaced here by 
\begin{align}
    \rho(\tau):= \frac{1}{2}T(\tau)-\tau\,.
    \label{eq: SW rho}
\end{align} 
Again, $\rho$ is positive and has a bounded limit $\rho_\infty$ for a family of SWs of \eqref{eq: DDE} where both $\tau$ and $T$ are tending to infinity. 
The system~\eqref{eq: APE1}, \eqref{eq: APE2}  with $\sigma=-\rho$ extends the properties of the profile equation~\cite{YanchukRuschelSieberWolfrum2019TDS} to SWs and we will refer to it as the {\it alternated profile equation (APE)}. 

Note that the APE-system \eqref{eq: APE1}, \eqref{eq: APE2} is equivariant with respect to the mirror symmetry
\begin{align}
    \gamma : (x, y) \mapsto (y, x)\,.
    \label{eq: mirror symmetry}
\end{align}
The subspace of fixed points of the symmetry  $\gamma$ is given by
\begin{align}
    \mathrm{Fix}(\gamma) := \lbrace (x,y)\;| \;\gamma(x, y)= (x,y) \rbrace = \lbrace(x,y)\;|\;  x=y \rbrace.
    \label{eq: symmetry subspace}
\end{align}
Obviously, the solutions in the symmetry subspace $\mathrm{Fix}(\gamma)$ correspond to solutions of the original system \eqref{eq: DDE} with $\tau=\sigma$.
The equilibria $(x_1,\, x_2)$ and $(x_2,\, x_1)$ given by the period-two orbit of the singular map are symmetry twins with respect to  $\gamma$ and are for  $\sigma =  -\rho$ both of saddle type. 
For $\rho \rightarrow \rho_\infty$ the period of the SW grows to infinity and the corresponding family of periodic solutions limits to a heteroclinic cycle, where the corresponding saddles are  $(x_1,\, x_2)$ and $(x_2,\, x_1)$. Note that also the two heteroclinics in this cycle are related by $\gamma$. Hence, it is sufficient to calculate one of them and we can describe the SW by a single connecting orbit $x_h (t)$, $x\in \mathbb{R}$ with 
\begin{align}\lim_{t\rightarrow \pm\infty}(x_h(t),y_h(t))= (x_{1,2}, x_{2,1})\,.\label{eq:het}\end{align}
This orbit  is connecting $(x_1,\, x_2)$ with its symmetry twin $(x_2,\, x_1)$ and can be called a {\em relative homoclinic} with respect to $\gamma$, since its source and target equilibrium are identical up to symmetry. All together, this allows  us to treat SWs in a similar manner as the TLSs described above.
As for TLSs, we can derive equation with an advanced argument, in which  they are given as a family of large-period periodic solutions that approach a relative homoclinic orbit corresponding to the limit of infinite delay.

Similar as in the profile equation for TLSs also in the APE-system~\eqref{eq: APE1}, \eqref{eq: APE2}, the response time $\rho$ appears as an additional parameter. This corresponds to the fact that connecting orbits satisfying Eq.~\eqref{eq:het} are objects of codimension one, i.e., they can be found only after adjusting one parameter. Hence, solving the APE for a homoclinic solution representing an SW, one needs to solve for the connecting orbit \eqref{eq:het} together with its response time $\rho_\infty$. In the original system \eqref{eq: DDE} the response time $\rho$ is determined according to Eq.~\eqref{eq: SW rho} as the difference of the delay and half of the period.
Studying bifurcations of TLSs induced by homoclinic bifurcations in the corresponding profile equation, one always has to also adjust the parameter $\rho_\infty$, which increases the codimension of the bifurcation by one. In the remaining sections of this paper, we  will use this approach to study the SWs in the DAE-system~\eqref{eq: KGTI1}, \eqref{eq: KGTI2}.

\section{\label{sec: SW in the APE}Finding square waves in the KGTI as connecting orbits}
Following the previous discussion, we introduce the alternated profile equation of the model for the Kerr-Gires-Tournois interferometer (KGTI) \eqref{eq: KGTI1}, \eqref{eq: KGTI2}  as
\begin{align}
\dot{E}&=\left[i\left(|E|^{2}-\delta\right)-1\right]E+hY, \label{eq: APE KGTI1}\\
Y&=\eta e^{i\varphi}\left(F\left(t+\rho\right)-Z\left(t+\rho\right)\right)+\sqrt{1-\eta^{2}}Y_{0},  \label{eq: APE KGTI2}\\
\dot{F}&=\left[i\left(|F|^{2}-\delta\right)-1\right]F+hZ,  \label{eq: APE KGTI3}\\
Z&=\eta e^{i\varphi}\left(E\left(t+\rho\right)-Y\left(t+\rho\right)\right)+\sqrt{1-\eta^{2}}Y_{0}\,.  \label{eq: APE KGTI4}
\end{align}
Note that the theory developed in Sec.~\ref{sec: SW as TDS} for delay-differential equations works similarly for a system of delay-algebraic equations, which in general are obtained by multiplying the left hand side of \eqref{eq: DDE} with a singular mass matrix $M$.
As before, we will study the resulting system using the continuation package DDE-BIFTOOL. 
It provides tools for a numerical computation of periodic solutions as boundary value problems with periodic boundary conditions. A path-continuation of periodic solutions can be performed by varying one parameter and adapting the period. For a continuation of a periodic solution with fixed period, yet another parameter has to be varied. Beyond that, the computation of connecting orbits is possible. They are implemented as a boundary value problem on a finite time domain, using boundary conditions projecting into the spectral eigenspaces of the corresponding saddle equilibria, see Refs.~\onlinecite{Beyn1990NumericalCompConnOrb, SamaeyEngelborghsRoose2002NumericalComputationConnectingOrb} for more details. Generic homoclinics are objects of codimension one, i.e. one has to solve for one parameter and  their continuation requires to treat two parameters as additional unknowns.
In many cases it is also possible to  approximate homoclinic orbits  by nearby periodic solutions with fixed large period. However, using spectral projection boundary conditions is often more robust and efficient. 

\begin{figure}
	\includegraphics[width=1\columnwidth]{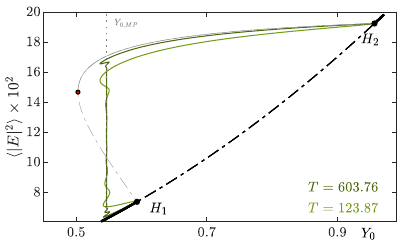}
	\caption{Black curve: branch of fixed points of the singular map of the DAE-system~\eqref{eq: KGTI1}, \eqref{eq: KGTI2}. Grey curve:  Branch of period-two orbits of the singular map bifurcating in flip bifurcations at $H_{1,2}$. Solid and dash-dotted parts indicate stable and unstable solutions, respectively. Green curves: branches of periodic solutions of the APE-system \eqref{eq: APE KGTI1}--\eqref{eq: APE KGTI4} with fixed periods $T=(603.76,123.87)$ and varying $\rho$. Other parameters cf. Fig. \ref{fig:fig2}.
 } \label{fig:fig3}
\end{figure}
Before we study the different types of periodic solutions and connecting orbits, we shortly describe the stationary solutions of the DAE-system~\eqref{eq: KGTI1}, \eqref{eq: KGTI2}. First, it has a branch of stationary solutions $x_{\mathrm{cw}}=(E_{\mathrm{cw}},Y_{\mathrm{cw}})$ representing CW states, see Fig.~\ref{fig:fig3}. They can also be found as fixed points of the corresponding singular map. These fixed points are unstable in a region between the two flip bifurcations located at $H_{1,2}$, where a branch of period-two orbits $(x_1,x_2)=((E_1,Y_1),(E_2,Y_2))$ of the singular map 
appears (grey line in Fig.~\ref{fig:fig3}). Note that in the regime shown in Fig.~\ref{fig:fig3} the flip bifurcation at the point $H_2$ is supercritical, while the other at $H_1$ is subcritical and the branch of period-two orbits has a fold (red point). In the APE-system \eqref{eq: APE KGTI1}--\eqref{eq: APE KGTI4} the CW state gives rise to an equilibrium solution $$(E,Y,F,Z)=(E_{\mathrm{cw}},Y_{\mathrm{cw}},E_{\mathrm{cw}},Y_{\mathrm{cw}}) \in \mathrm{Fix}(\gamma),$$ while the  period-two orbit of the singular map corresponds to symmetry related equilibrium solutions 
$$ (E,Y,F,Z)=(E_1,Y_1,E_2,Y_2),\quad (E,Y,F,Z)=(E_2,Y_2,E_1,Y_1) .$$
As pointed out above, their stability properties in the APE differ from the stability in the DAE-system \eqref{eq: KGTI1}, \eqref{eq: KGTI2}. Indeed, all these equilibria of the APE-system \eqref{eq: APE KGTI1}--\eqref{eq: APE KGTI4} are of saddle type with two complex conjugate pairs of leading eigenvalues (double focus). 
Close to the flip bifurcations of the singular map at $H_{1,2}$ we find in the APE-system AH-bifurcations. Fixing the period $T$ and solving simultaneously for the corresponding value of $\rho$, we obtain the emanating branches of different periodic solutions (cf. green curves in Fig.~\ref{fig:fig3}). 

\begin{figure}
	\includegraphics[width=1\columnwidth]{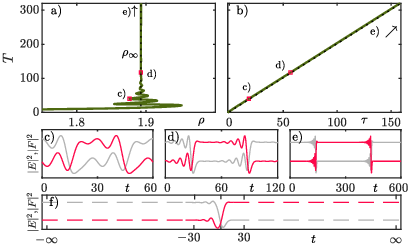}
	\caption{Branches of SWs in the (a) APE-system \eqref{eq: APE KGTI1}--\eqref{eq: APE KGTI4}, and (b) DAE-system \eqref{eq: KGTI1}, \eqref{eq: KGTI2}. In (b) the branch approaches the line $2\tau + 2\rho_\infty$ (dashed), for $\tau,T \rightarrow \infty$. In (a) the branch approaches a vertical line (dashed) at $\rho=\rho_\infty$ with $T \rightarrow \infty$ in a homoclinic bifurcation. (c)--(e) show profiles for the periods $T \approx 60,120,600$, respectively. For $\rho \rightarrow \rho_\infty$ the SWs approximate the relative homoclinic given in (f). Parameters as in Fig. \ref{fig:fig2} and $Y_0=0.7$.
 } \label{fig:fig4}
 \end{figure}

\begin{figure}
	\includegraphics[width=1\columnwidth]{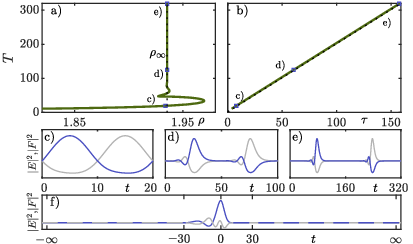}
	\caption{Branches of alternating pulse solutions in the (a) APE-system \eqref{eq: APE KGTI1}--\eqref{eq: APE KGTI4}, and (b) DAE-system \eqref{eq: KGTI1}, \eqref{eq: KGTI2}. In (b) the branch approaches the line $2\tau + 2\rho_\infty$ (dashed), for $\tau,T \rightarrow \infty$. In (a) the branch approaches a vertical line (dashed) at $\rho=\rho_\infty$ with $T \rightarrow \infty$ in a homoclinic bifurcation. (c)--(e) show profiles for the periods $T = 20,100,320$. For $\rho \rightarrow \rho_\infty$ the periodic solutions approximate the homoclinic given in (f). Parameters as in Fig. \ref{fig:fig2} and $Y_0=0.58$.
 } \label{fig:fig5}
 \end{figure}

Close to the point $H_2$, the branches of periodic solutions have the shape of SWs and follow the stable branch of the period-two orbit of the singular map, approaching it for large $T$.
The corresponding SW solutions of system~\eqref{eq: KGTI1}, \eqref{eq: KGTI2} inherit the stability from the period-two orbit of the singular map. This behavior close to a supercritical flip bifurcation of the singular map has been studied extensively from a mathematical point of view, see e.g. Refs.~\onlinecite{MalletParet1986GlobalCA,HaleHuang1994PeriodDoubling, HaleHuang1996singularlyperturbedDDE}. 
Figure \ref{fig:fig4} illustrates the SW solutions, which can be found numerically for a fixed value of $Y_0$ close to $H_2$. Panel (a) shows a branch of SW solutions in the APE-system \eqref{eq: APE KGTI1}--\eqref{eq: APE KGTI4} with varying parameter $\rho$ and resulting period $T$. This branch of periodic solutions of the APE limits to a relative homoclinic orbit \eqref{eq:het} with $\rho \rightarrow \rho_\infty$ and $T\rightarrow \infty$. The branch reappears in the DAE-system \eqref{eq: KGTI1}, \eqref{eq: KGTI2} as a branch of stable SW solutions with $T, \tau\rightarrow \infty$, as shown in panel (b). Panels (c)--(e) show the intensity profiles of selected periodic solutions for the fields $E$ (red) and $F$ (grey) of the APE-system along this branch with increasing period. Panel (f) depicts a numerical representation of the limiting connecting orbit \eqref{eq:het}, where we employed spectral projection boundary conditions to the stable and unstable manifolds of the saddle equilibria $(x_{1,2},x_{2,1})$ at the endpoints of a sufficiently long computational interval. The snaking shape of the branch of periodic solutions in the panel (a) is a consequence of the complex conjugated leading eigenvalues at the saddle equilibria $(x_{1,2},x_{2,1})$, which induce also the oscillatory tails of the localized structure in the panel (f), see e.g., Ref.~\onlinecite{ShilnikovShilnikovTuraevChua2001NonlinearDynamics2} for details. Note that the frequency of these damped oscillations in the tails remains unchanged for the different values of $T$ and the corresponding delay times $\tau$ such that the plateaus of different lengths of the SW solutions in Fig.~\ref{fig:fig4} (c)--(f) carry different numbers of such oscillations. 
Close to the bifurcation point $H_1$, the period-two orbit of the singular map is unstable and the branches of periodic solutions follow instead the stable branch of CW states and approach it for large $T$. The corresponding solutions in the DAE-system \eqref{eq: KGTI1}, \eqref{eq: KGTI2} are unstable alternating pulse solutions, 
which spend most of the time between the pulses close to the stable CW state.
The emergence of such solutions close to a subcritical flip bifurcation of the singular map has also been studied from a mathematical point of view~\cite{HaleHuang1994PeriodDoubling, HaleHuang1996singularlyperturbedDDE}. Figure \ref{fig:fig5} illustrates these solutions, which can be found numerically for a fixed value of $Y_0$ close to $H_1$. Panel (a) shows a branch of alternating pulse solutions in the APE-system \eqref{eq: APE KGTI1}--\eqref{eq: APE KGTI4} with varying parameter $\rho$ and resulting period $T$, limiting to a homoclinic orbit with $\rho \rightarrow \rho_\infty$ and $T\rightarrow \infty$. Again this branch reappears in the DAE-system \eqref{eq: KGTI1}, \eqref{eq: KGTI2}  with increasing $\tau\rightarrow \infty$ as depicted in panel (b). Panels (c)--(e) represent the intensity profiles of the fields $E$ (blue) and $F$ (grey) from \eqref{eq: APE KGTI1}--\eqref{eq: APE KGTI4} of selected periodic solutions, while panel (f) demonstrates the limiting orbit, which is a homoclinc to  $(x_{\mathrm{cw}},x_{\mathrm{cw}})$. Note that the limiting saddle equilibrium of this homoclinic orbit is in the symmetry subspace $\mathrm{Fix}(\gamma)$ while the homoclinc orbit itself is not: The pulse shapes in the two components of the APE are different, which leads to the alternating behavior of  the corresponding periodic solutions in the DAE-system \eqref{eq: KGTI1}, \eqref{eq: KGTI2}.

Figure~\ref{fig:fig3} shows that the branches of periodic SW solutions emanating from $H_2$ follow the branch of stable period-two solutions of the singular map only up to a certain region where they detach and enter a complicated snaking structure. In the same parameter region also the branches of alternating pulse solutions emanating from $H_1$ detach from the  CW states and  enter the same snaking structure from the other side. For larger delay and correspondingly larger periods the number of folds in this snaking region increases. This leads to multiple coexisting stable solutions of mixed type, i.e. a combination of square waves and alternating pulses.
We will study this dynamical phenomenon in detail in the following section.

\section{\label{sec: T-pt}Appearance of mixed type solutions through a Bykov T-point} 
\begin{figure}
	\includegraphics[width=1\columnwidth]{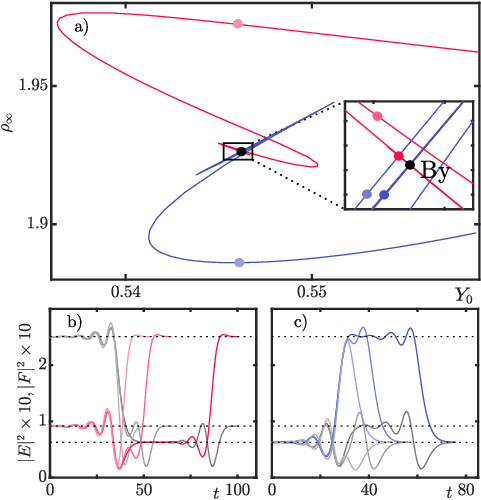}
	\caption{Branches of connecting orbits for the APE-system \eqref{eq: APE KGTI1}--\eqref{eq: APE KGTI4} with varying parameters $Y_0$ and $\rho_\infty$. Red curve in (a): relative homoclinic orbits connecting the equilibria $(x_{1,2},x_{2,1})$ corresponding to the period-two orbit of the singular map. Panel (b) shows selected profiles (red points in (a)) from this branch (red: $|E(t)|^2$, grey: $|F(t)|^2$). Blue curve in (a): homoclinic orbits to the equilibrium $(x_{\mathrm{cw}},x_{\mathrm{cw}})$ corresponding to the CW state. Panel (c) shows selected profiles from this branch (blue points in (a)). Both branches are spiraling into a Bykov T-point (black dot).
    Other parameters as in Fig.~\ref{fig:fig2}.
    } \label{fig:fig6}
\end{figure}

We will now use the limiting (relative) homoclinic orbits introduced before as an approximation of their nearby periodic solutions, which, at the same time, provide a qualitative understanding of the shape of their profiles and how their branches are organized in parameter space. 
Figure~\ref{fig:fig6} shows the two branches of solutions starting from the points $H_{1,2}$, respectively, and entering into the snaking region. They are now calculated  as connecting orbits for the APE-system \eqref{eq: APE KGTI1}--\eqref{eq: APE KGTI4} with varying parameters $Y_0$ and $\rho_\infty$.  Panel (a) in Fig.~\ref{fig:fig6} shows only a small region in $Y_0$ chosen in the vicinity of the snaking region (compare the range of  $Y_0$  in Fig.~\ref{fig:fig3}). 

The solutions on the red branch are relative homoclinic orbits \eqref{eq:het} with limiting equilibria $(x_{1,2},x_{2,1})$, which correspond to the period-two orbit of the singular map. In the representation in the $(Y_0,\,\rho_\infty)$-plane the branch  when entering the snaking region starts spiraling and ends after infinitely many folds at the point $(Y_0^\star,\rho_\infty^\star)\approx (0.546,1.926)$ (black dot). 
Panel (b) shows the intensity profiles of $E$ (red) and $F$ (grey) fields of selected solutions along the red branch (cf. corresponding points in shades of red in Fig.~\ref{fig:fig6}~(a)). Note that the numerically  calculated profiles extend over time intervals of different lengths. This results from the fact that the solutions are calculated only outside a certain distance from the limiting equilibria. For the different solutions it takes different time spans to pass over this distance.
Before the first fold on the red branch the solutions are of SW type corresponding to stable solutions in the DAE-system \eqref{eq: KGTI1}, \eqref{eq: KGTI2}. After the first fold they become unstable, but restabilize after a second fold, and we obtain solutions that do not directly switch from  $(x_{1},x_{2})$ to $(x_{2},x_{1})$ but in between perform a single oscillation around the CW state $(x_{\mathrm{cw}},x_{\mathrm{cw}})$ in the corresponding APE-system. All such solutions coexist as stable solutions of the DAE-system \eqref{eq: KGTI1}, \eqref{eq: KGTI2} with the solutions of SW type before the first fold.  Moving further along the spiraling branch we get further solutions showing an increasing number of damped oscillations around the CW state and spending a longer time there. At the same time the interval of parameter $Y_0$ where they exist gets smaller and smaller. These mixed-type solutions are connecting orbits from $(x_{1},x_{2})$ to $(x_{2},x_{1})$ that develop a plateau of increasing length at the CW state. 

The blue branch of homoclinic orbits to the equilibrium $(x_{\mathrm{cw}},\,x_{\mathrm{cw}})$ shown in Fig.~\ref{fig:fig6}~(a) has a similar spiraling shape and also ends after infinitely many folds at the point $(Y_0^\star,\rho_\infty^\star)$. The solutions along this branch behave in a complementary manner, see Fig.~\ref{fig:fig6}~(c). Before the first fold they provide alternating pulse solutions, which correspond to unstable solutions in the DAE-system~\eqref{eq: KGTI1}, \eqref{eq: KGTI2}. Then, after the first fold, they become stable. This happens when the two  pulses come close to the levels $x_1$ and $x_2$ of the stable period-two orbit of the singular map and the solution stays  there for some time. Moving further into the spiral we get further stable solutions showing an increasing number of damped oscillations  around the levels of the SW and spending a longer time there, i.e., we obtain mixed-type solutions being homoclinic orbits to $(x_{\mathrm{cw}},x_{\mathrm{cw}})$  that develop a plateau of increasing length at the equilibrium  $(x_{1},x_{2})$. Again, with the increasing length of the plateau the parameter intervals of their existence becomes smaller.  
\begin{figure}
	\includegraphics[width=0.9\columnwidth]{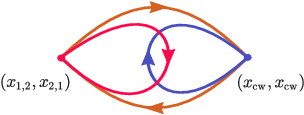}
	\caption{Sketch of connecting orbits involved in the unfolding of a Bykov T-point. A homoclinic switches its saddle by transitioning through a heteroclinic cycle (orange). The homoclinic solutions involved in the unfolding are a relative homoclinic to $(x_{1,2},x_{2,1})$ (red) and to  $(x_{\mathrm{cw}},x_{\mathrm{cw}})$
 (purple), representing the SW solutions and alternating pulse solutions respectively.} \label{fig7}
\end{figure}

At the endpoint $(Y_0^\star,\rho_\infty^\star)$ of the two branches we find a so-called Bykov T-point \cite{Bykov1993Tpts,GlendinningSparrow1986Tpoints,KnoblochLamb2014Bykov}. Such points arise as homoclinic bifurcations of codimension two when a  homoclinic orbit collides with another saddle equilibrium. In this way we obtain a family of homoclinic orbits that switches at the T-point its limiting saddle to another saddle.
A schematic representation of this procedure is shown in Fig.~\ref{fig7}. The bifurcation point itself is characterized as a heteroclinic cycle connecting two different saddle equilibria (orange arrows). In a neighborhood in the parameter space there can be found  homoclinic orbits to each of the two equilibria (cf. red and blue lines in Fig.~\ref{fig7}). 
Note that the situation in the APE-system \eqref{eq: APE KGTI1}--\eqref{eq: APE KGTI4} fits to this general scenario when we identify the equilibrium  $(x_{1},x_{2})$  with its symmetry twin  $(x_{2},x_{1})$. Such Bykov T-points can have different types depending on the leading eigenvalues of the saddles being real or complex conjugate pairs. Figure~\ref{fig7} depicts the situation with only real leading eigenvalues, which actually can be realized for a planar ordinary differential equation. For the APE-system here, both saddle equilibria $(x_{\mathrm{cw}},x_{\mathrm{cw}})$ and  $(x_{1},x_{2})$  are of double focus type i.e, having two complex conjugate pairs as leading eigenvalues. This leads to the spiraling shape of the two branches in Fig.~\ref{fig:fig6}~(a). In all cases there are two branches of homoclinic orbits emanating from the T-point, each of them to one of the two equilibria. 

\begin{figure}
	\includegraphics[width=1\columnwidth]{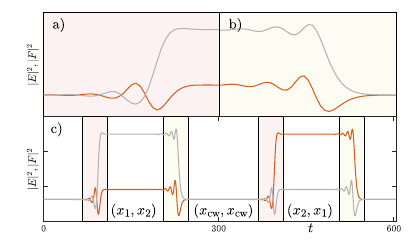}
	\caption{Profiles of the two heteroclinic solutions at the Bykov T-point: (a)--heteroclinic solution connecting $(x_\mathrm{cw},x_\mathrm{cw})$ to $(x_1,x_2)$; (b)--heteroclinic solution connecting $(x_1,x_2)$ to $(x_\mathrm{cw},x_\mathrm{cw})$; (c)--profile of a periodic solution of mixed type with $T=600$, composed by the heteroclinc solutions and plateaus at the equilibria. Intensity profiles of the fields $E$ (orange) and $F$ (grey) are shown. Parameters as in Fig.~\ref{fig:fig2}.} \label{fig:fig8}
\end{figure}

In Ref.~\onlinecite{StohrWolfrum2023TDS} it has been shown that branches of stable TLSs can terminate at Bykov T-points where the corresponding homoclinic orbit of the profile equation collides with another equilibrium. Indeed, when the second equilibrium is unstable in the original DDE the T-point leads to a delocalization and destabilization of the solution. Here, both equilibria involved in the T-point are stable in the singular map. Hence, in the DAE-system \eqref{eq: KGTI1}, \eqref{eq: KGTI2} the collision of the SW with the stable CW state and, likewise, the collision of the alternating pulses with the stable period-two orbit leads to stable solutions of mixed type. They are given as the periodic orbits in the vicinity of the branches of homoclinic orbits ending at the T-Point in the corresponding APE-system. 
Indeed, it is known that in the vicinity of a T-point there are periodic orbits that can spend arbitrarily long times close to each saddle equilibrium. In Fig.~\ref{fig:fig8} we demonstrate how the profiles of these periodic orbits corresponding to the different mixed-type solutions can be put together using the heteroclinic solutions at the T-point.  
Panels (a) and (b) depict the two heteroclinic solutions connecting $(x_{\mathrm{cw}},x_{\mathrm{cw}})$ 
and $(x_1,x_2)$ and vice versa. A periodic solution with $T \approx 600$ close to the heteroclinic cycle for the same parameters is given in Fig.~\ref{fig:fig8}~(c). Such a periodic solution can be well-approximated by using the segments of the heteroclinic solutions from the panels (a) and (b) at the transition layers (red and yellow shaded regions, respectively) and in-between them inserting plateaus at $(x_{\mathrm{cw}},x_{\mathrm{cw}})$ and  $(x_{1,2},x_{2,1})$, which can have arbitrary length (white regions).

The snaking branches in the vicinity of the Bykov T-point can be found numerically both by path-continuation of periodic solutions with large fixed period, which are close to the connecting orbit, or by continuation of the connecting orbit itself. The continuation of the T-point itself is only possible as a connecting orbit. 
In general, a Bykov T-point is a bifurcation of codimension two. In the APE-system~\eqref{eq: APE KGTI1}--\eqref{eq: APE KGTI4} both saddle equilibria  have the same saddle index, i.e. the same number of unstable eigenvalues. Hence, a single heteroclinic solution connecting them is of codimension one. For the T-point we have to find parameter values where both heteroclinic connections, back and forth,  exist simultaneously, which implies the codimension two. For a continuation of such a codimension-two object one has to vary three parameters. However, in the APE-system \eqref{eq: APE KGTI1}--\eqref{eq: APE KGTI4} one has the additional parameter $\rho$, which is absent in the DAE-system \eqref{eq: KGTI1}, \eqref{eq: KGTI2}. Hence, we can represent the curve of Bykov T-points in the parameter plane $(Y_0,\delta)$, keeping in mind that at each point on the branch our calculations have provided also a response time $\rho_\infty$. 

\begin{figure}
	\includegraphics[width=1\columnwidth]{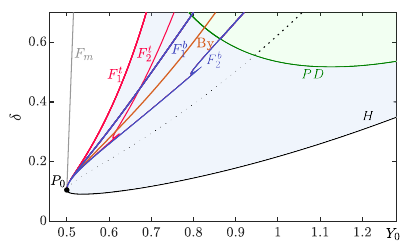}
	\caption{Bifurcation diagram for system \eqref{eq: KGTI1}, \eqref{eq: KGTI2} in the $(Y_0,\delta)$-plane: continuation of the Bykov T-point (orange),  branch of supercritical (solid black) and subcritical (dotted black) AH-points, first two folds along the branch of SWs (red), first two folds along the branch of alternating pulse solutions (blue). Fold of the equilibrium period-two orbit of the singular map (grey). All lines terminate in the point $P_0$ of the branch of AH-points. Green line corresponds to the onset of the period-doubling bifurcation of SWs. Other parameters as in Fig.~\ref{fig:fig2}.
 } \label{fig:fig9}
\end{figure}
Figure~\ref{fig:fig9} shows a bifurcation diagram of the system \eqref{eq: KGTI1}, \eqref{eq: KGTI2}, cf. Fig. 3 in Ref.~\onlinecite{KSG-OL-22}. It shows the curve of Bykov T-points (orange) in the $(Y_0,\delta)$ plane together with the AH-bifurcations of the CW state (solid and dotted black for supercritical and subcritical points, respectively), the fold of the period-two orbit of the singular map (grey), and the two first fold bifurcations along the branch of SWs (red) and alternating pulse solutions (blue). Note, that the position of the Bykov T-point provides the position of the Maxwell point at the center of the snaking region. The first fold of the SWs $F_{1}^t$ together with the AH-line $H$ provides the border of the region of stable SW solutions (blue shaded region). The other border is given by a period-doubling of the singular map (green line $PD$), which appears for larger values of $\delta$ and $Y_0$. 

Between the fold of the period-two orbit of the singular map  $(x_1,x_2)$, which is also a fold of the corresponding equilibrium of the APE-system (grey line $F_m$), and the fold  $F_{1}^t$ of the SW there is a gap region with stable period-two orbits of the singular map but no stable SW solutions. This is related to the fact that the stable period-two orbit coexists with the stable CW state, which obstructs in this region from passing from one level of the period-two orbit to the other. The curve of T-points necessarily has to lie in the region of bistability of the singular map, localized between the curve $F^m$ and the subcritical part of the AH-curve (black dotted). Note that all the bifurcation curves emanate from  the  point $P_0\approx (0.500, 0.105)$. This is where the flip bifurcation of the singular map changes from sub- to supercritical. In the APE-system this corresponds to some type of degenerate Bykov T-point -- a codimension three homoclinic bifurcation, which we are not going to study in more detail in this work. For values of $\delta$ below the point $P_0$ the scenario for varying $Y_0$ is much simpler. There are two supercritical Hopf bifurcations and between them a branch of stable SWs without any bistability.

\section{Multiple T-points}
We started our study with a single branch of periodic solutions, which connects the two AH-bifurcations $H_{1,2}$ as shown in Fig.~\ref{fig:fig2}. Calculating the corresponding homoclinic orbits in the APE-system, we found that along this branch there is a single Bykov T-point, which explains the snaking behavior of the branch and the emergence of mixed-type solutions in the snaking region. However, it is known that saddle-focus or double-focus homoclinic orbits, i.e., with complex conjugate leading eigenvalues, under certain conditions can induce in their neighborhood in phase space and for nearby parameter values an extremely complicated structure of chaotic motion, including further periodic orbits and so-called $N$-homoclinic orbits, see Refs. \onlinecite{ShilnikovShilnikovTuraevChua2001NonlinearDynamics2,GlendinngSparrow1984HomoclinicOrbits}. For T-points involving saddles of saddle-focus or double-focus type it is also known that their unfolding 
can induce infinitely many other T-points involving infinitely many $N$-homoclinics for all $N \in \mathbf{N}$. So far, we did not pay attention to these additional solutions since the corresponding solutions of the original system  \eqref{eq: KGTI1}, \eqref{eq: KGTI2} all turn out to be unstable. 

Another feature of the given system \eqref{eq: APE KGTI1}--\eqref{eq: APE KGTI4} is that 
in the region of bistability of the CW state and the period-two orbit $(x_1,x_2)$ of the singular map there are also additional equilibria of system \eqref{eq: APE KGTI1}--\eqref{eq: APE KGTI4}, which correspond to the unstable branch of  period-two orbits of the singular map. These equilibria can also be involved in homoclinic orbits and T-points. In the case of an unstable  background, given by such an unstable period-two orbit, an SW has to be unstable as well. But there can be T-points where one of the saddle equilibria corresponds to a stable background, while the other saddle corresponds to an unstable background (see also Ref.~\onlinecite{StohrWolfrum2023TDS}). We will show now that for larger values of the detuning parameter $\delta$ the additional branches of homoclinic orbits and further T-points, involving also the unstable period-two orbit, can come into the play.

\begin{figure}
\includegraphics[width=1\columnwidth]{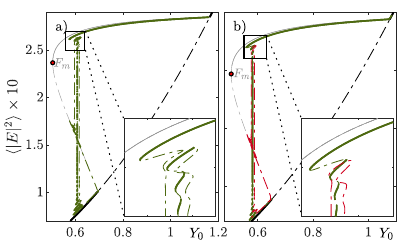}
\caption{a) For $\delta=0.29$ the branches of periodic solutions have reconnected, creating  a complex structure of multiple T-points along a single branch. b) For the lower value  $\delta=0.25$ this structure is detached and fully unstable (red). It contains  another unstable SW regime which has  the unstable period-two orbit of the singular map as its background.  Other parameters  as in Fig.~\ref{fig:fig2}. } \label{fig:fig10}
\end{figure}


Figure~\ref{fig:fig10}~(a) shows an exemplary SW branch calculated from the DAE-system~\eqref{eq: KGTI1}, \eqref{eq: KGTI2} for $\delta=0.29$. In contrast to the scenarios studied in the preceding sections, this branch features four T-points, one of them lies near the AH-bifurcation point and is not visible in the figure here. 
Along this branch we find also unstable SWs with an unstable background given by the unstable period-two orbit of the singular map (dash-dotted grey curve in Fig.~\ref{fig:fig10}) as predicted analytically in Ref.~\onlinecite{KSG-OL-22}. 
Moreover, two of the four T-points involve equilibria corresponding to such unstable backgrounds.
Among the remaining two T-points that involve the two stable backgrounds there is only one  
(middle curve of the three vertical green snaking curves in  Fig.~\ref{fig:fig10} (a)) that displays also stable regions, whereas the snaking branch from the other such T-point caries only unstable solutions (see the inset in Fig.~\ref{fig:fig10} (a)).

Decreasing the detuning parameter $\delta$ one can observe a recombination of the branches (see the inset of Fig.~\ref{fig:fig10} (b)). As a result we obtain two separate branches, see Fig.~\ref{fig:fig10}~(b). The red branch is now a closed loop not connected to $H_{1,2}$ and carries  only unstable solutions and  T-points involving the unstable background. It is detached from the primary branch (green) connecting to $H_{1,2}$ and carrying the stable SWs, the single T-point, and the stable mixed-type solutions. Note that this reconnection of the branches does not substantially change the scenario of stable solutions. There is still a branch of stable SWs close to the stable period-two orbit of the singular map, which ends close to the Maxwell point indicated by the primary Bykov T-point.  In the vicinity of this T-point we find the nested intervals of stable mixed-type solutions. For the continuation of the T-point it is not relevant whether the emanating branches of homoclinics connect directly to the AH-points $H_{1,2}$. Continuating the primary T-point as in Fig.~\ref{fig:fig9}, we stay with the branches carrying the stable mixed-type solutions also for higher values of $\delta$.


%

\section{Conclusion}
In conclusion, in this paper we unveiled the mechanisms responsible for the formation of collapsed snaking of square waves in vertical external-cavity Kerr-Gires–Tournois interferometers from the point of view of homoclinic bifurcation theory. Starting from  a delay algebraic equations model in the long delay limit, we introduced the so called {\em alternated profile equation}, which allows to find the SW's time profiles. In this equation the large delay has disappeared and SWs with arbitrarily high periods can be found close to connecting orbits at finite parameter values. These connecting orbits can be seen as relative homoclinic orbits with respect to a mirror symmetry. In this way, we can exploit the limit of large delay to obtain  
a qualitative understanding of the shape of the profiles and how their branches are organized in parameter space. In particular, we demonstrated that the position of the limiting Bykov T-point, arising as homoclinic bifurcations of codimension two when a homoclinic orbit collides with another saddle equilibrium, corresponds to the position of the Maxwell point in the center of the snaking region.

This introduces a new tool in the analysis of temporally localized solutions in time-delayed systems and their  bifurcations.  Certain types of T-points  can indicate  a region with collapsed snaking, which can induce multistability of periodic solutions with  complicated multiple localized structures. 
This approach can be applied to explain a wide range of different scenarios. An example is the observation of coexistence regions of dark and bright temporal localized states in the DAE-system~\eqref{eq: KGTI1}, \eqref{eq: KGTI2} that evolve in a collapsed snaking scenario around a Maxwell point as shown in Refs.~\onlinecite{SPV-OL-19,SJG-OL-22,SGJ-PRL-22}. Additionally, the derivation of an alternated profile equation for periodic alternating plateau solutions with a period close to twice the delay could be extended to $N\tau$ periodic solutions with multiple plateaus. Especially for the second example the results of this work could possibly help to reveal and to understand the involved complex bifurcation scenarios.
%
%
Finally, the alternated profile equation can also be used to calculate period-doublings of the periodic soluions of time-delayed systems and to find branches of period-doubled solutions.  This is out of the scope of the current paper and is a subject of the ongoing work.

\section*{Author Declarations}
The authors have no conflicts to disclose. $^{\dagger}$ These authors contributed equally to this paper.

\section*{Acknowledgments}
E.K, J. J. and S.G. acknowledge the financial support of the projects KEFIR, Ministerio de Economía y Competitividad (PID2021-128910NB-100 AEI/FEDER UE) and KOGIT, Agence Nationale de la Recherche (ANR-22-CE92-0009)), Deutsche Forschungsgemeinschaft (DFG) via Grant Nr. 505936983.

\section*{Data Availability}
The data that support the findings of this study are available from the corresponding author upon reasonable request.

\end{document}